\documentclass[aps,prb,twocolumn,showpacs]{revtex4}
\usepackage{latexsym}
\usepackage{amssymb}
\usepackage{bm}
\usepackage{graphicx}

\newcommand{\pdag}{{\phantom{\dagger}}}
\newcommand{\bq}{\begin{equation}}
\newcommand{\eq}{\end{equation}}
\newcommand{\bn}{\begin{eqnarray}}
\newcommand{\en}{\end{eqnarray}}

\begin{document}

\title{Time-dependent resonant tunneling for a parallel-coupled double quantum dots}

\author{Bing Dong$^{1,2}$, Ivana Djuric$^{1}$, H. L. Cui$^{1,3}$, and X. L. Lei$^{2}$} 
\affiliation{$^{1}$Department of Physics and Engineering Physics, Stevens Institute of 
Technology, Hoboken, New Jersey 07030 \\
$^{2}$Department of Physics, Shanghai Jiaotong University,
1954 Huashan Road, Shanghai 200030, China \\
$^{3}$School of Optoelectronics Information Science and Technology, Yantai University, 
Yantai, Shandong, China}

\begin{abstract}

We derive the quantum rate equations for an Aharonov-Bohm interferometer with two 
vertically coupled quantum dots embedded in each of two arms by means of the 
nonequilibrium Green's function in the sequential tunneling regime. Basing on these 
equations, we investigate time-dependent resonant tunneling under a small amplitude 
irradiation and find that the resonant photon-assisted tunneling peaks in photocurrent 
demonstrate a combination behavior of Fano and Lorentzian resonances due to the 
interference effect between the two pathways in this parallel configuration, which is 
controllable by threading the magnetic flux inside this device.   

\end{abstract}


\maketitle

\section{Introduction}

The investigation of quantum coherence in mesoscopic systems has been the subject of 
considerable interest in the solid state physics during the last years. In the 
interference experiments with a quantum dot (QD) embedded in one arm of the 
Aharonov-Bohm (AB) ring, the tunneling through a QD was proved to be coherent by 
detecting the flux-periodic current oscillations.\cite{Hackenbroich} More recently, 
Holleitner {\it et al}. extended this idea to measure the AB oscillations of the 
mesoscopic ring containing two coupled QDs inserted in each of the two 
arms.\cite{Holleitner1} Furthermore, this parallel-coupled QDs structure has been 
investigated in the Kondo regime, and an observation of the transition between different 
quantum states has been reported.\cite{Holleitner2,Chen} Clearly, the possibility to 
manipulate each of the QDs separately and the application of the magnetic flux provide 
more controllable parameters for designing the transport properties. This has been 
discussed by several theoretical works for the stationary transport by nonequilibrium 
Green's function (GF).\cite{Loss,Konig,Kubala,Shahbazyan,Ladron}

On the other hand, time-dependent tunneling through coupled QDs in series has received 
large attention both theoretically and experimentally. A theoretical study of the 
photon-assisted tunneling (PAT) in double QDs given by Stoof and Nazarov\cite{Stoof} and 
Hazelzet {\it et al}.\cite{Hazelzet}, based on the quantum rate equation approach, 
predicted that the photoresponse of the system exhibits satellite resonance peaks due to 
PAT processes which involved the emission or absorption of one photon to match the 
energy difference between the discrete states of the two QDs. W. G. van der Wiel {\it et 
al}.\cite{Wiel} measured the PAT current through weakly coupled QDs and discovered 
clearly the predicted extra resonance peaks under irradiation of microwave. Motivated by 
this perfect match between the theory and experiments, we intended, in this paper, to 
study the PAT in the parallel-coupled QDs by the quantum rate equations approach. In 
this configuration, the additional bridges between the QDs and leads allow the electron 
wavefunction propagating along different pathways, then lead to the interference effect 
between them, which is displayed by the fundamental AB oscillation in the presence of a 
magnetic field. Therefore, the central point of our study is to explore how the 
interference influences the photoresponse of the parallel-coupled QDs.

The rest of the paper is organized as follows. First, in the following section, we 
establish the quantum rate equations for this system in the presence of a magnetic field 
by employing the nonequilibrium Green's function.\cite{Dong,Ma} Then in the section III, 
we calculate the current as a function of magnetic fluxes, and study the quantum 
dynamics of this system. The spectrum investigation in Ref.\onlinecite{Ladron} pointed 
out that increasing strength of the additional bridges causes the total localization of 
the antibonding state due to the perfect destructive interference, and as a consequence 
the transport characteristic of the device reduces approximately to a single QD. In this 
paper we restrict our interest to the regime where there are two distinctly resolved 
peaks in the density of states spectra and both of the bonding and antibonding states 
can contribute to transport. Our numerical results show that the current in this regime 
still keeps the oscillation behavior with magnetic flux but the period changes from 
$2\pi$ to $4\pi$ due to the interdot coupling. The temporal investigation of the 
electron occupation probabilities in the two QDs shows that the conventional oscillation 
behavior in a two-level system can be destroyed by the additional bridges connecting the 
two QDs and two leads, and it can be recovered by applying a nonzero magnetic flux. Also 
in this section we study in detail the photoresponse of the system subject to a small 
irradiation and predict novel enclosed magnetic flux-controlled photon-assisted peaks in 
tunneling current, which can be attributed to the interference between the two pathways 
electrons going through in this system. Finally, a summary is given in Section IV.       
       
\section{Model and formulation}

We consider the parallel-coupled interacting QDs interferometer connected to two normal 
leads as depicted in Fig.~1. Only one bare energy level in each dot is involved in 
transport. The intradot electron-electron Coulomb interactions are assumed to be 
infinite but the interdot interaction $U$ is finite. Namely, the state of two electrons 
occupied in the same QD is forbidden but two electrons dwelling in different QDs is 
permitted. 

\begin{figure}[htb]
\includegraphics[height=1.5in,width=2.5in]{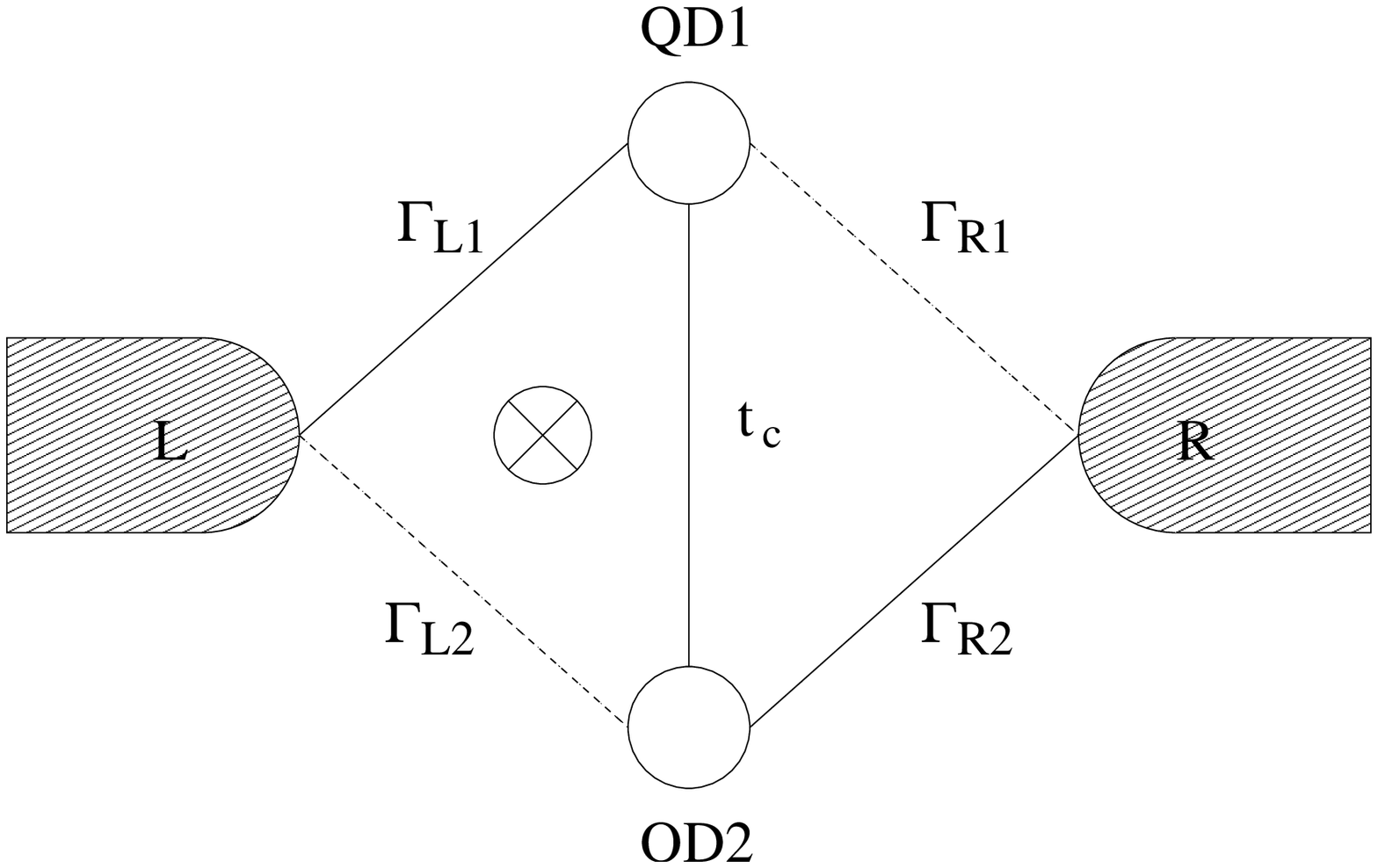}
\caption{Parallel-coupled quantum dots Aharonov-Bohm interferometer.}
\label{fig1}
\end{figure}

For the sake of simplicity, we abandon the spin because transport through this system is 
spin independent. Therefore, the available states and the corresponding energies for the 
interferometer with two embedded QDs are: (1) the whole system is empty, $|0\rangle_{1} 
|0\rangle_{2} $, and the energy is zero; (2) the first QD is singly occupied, 
$|1\rangle_{1} |0\rangle_{2}$, and the energy is $\epsilon_{1}$; (3) the second QD is 
singly occupied, $|0\rangle_{1} |1\rangle_{2}$, and the energy is $\epsilon_2$; (4) both 
of the QDs are singly occupied, $|1\rangle_{1} |1\rangle_{2}$, and the energy is 
$\epsilon_{1} +\epsilon_{2} + U$. We assign these Dirac brackets as operators: the 
slave-boson operators $e^{\dagger}=|0\rangle_{1} |0\rangle_{2}$, 
$d^{\dagger}=|1\rangle_{1} |1\rangle_{2}$ and the pseudo-fermion operators 
$f_{1}^{\dagger}=|1\rangle_{1} |0\rangle_{2}$, $f_{2}^{\dagger}=|0\rangle_{1} 
|1\rangle_{2}$. And obviously the explicit (anti)communicators of these auxiliary 
particles are:\cite{Dong}
\[
ee^{\dagger}=1, \quad d d^{\dagger} = 1, \quad f_{i}^{\pdag} 
f_{j}^{\dagger}=\delta_{ij},
\]
\bq
ed^{\dagger} = e f_{j}^{\dagger} = f_{j}^{\pdag}e^{\dagger} = f_{j}^{\pdag} d^{\dagger} 
= d e^{\dagger} = d f_{j}^{\dagger}=0, \label{quan}
\eq
in association with the completeness relation
\bq
e^{\dagger} e + d^{\dagger} d + f_{1}^{\dagger} f_{1 }^{\pdag} + f_{2}^{\dagger} 
f_{2}^{\pdag} =1.
\eq

The density matrix elements are expressed as $\rho_{00}=|0\rangle_{1} |0 \rangle_{2} 
{}_{2}\langle 0| {}_{1}\langle 0|=e^{\dagger} e$, $\rho_{11}=|1\rangle_{1} |0 
\rangle_{2} {}_{2}\langle 0| {}_{1}\langle 1| =f_{1}^{\dagger} f_{1}^{\pdag}$, 
$\rho_{22}=|0\rangle_{1} |1 \rangle_{2} {}_{2}\langle 1 | {}_{1}\langle 
0|=f_{2}^{\dagger} f_{2}^{\pdag}$, $\rho_{dd}=|1\rangle_{1} |1 \rangle_{2} {}_{2}\langle 
1| {}_{1}\langle 1|=d^{\dagger} d$, and $\rho_{12}=|0\rangle_{1} |1 \rangle_{2} 
{}_{2}\langle 0| {}_{1}\langle 1|= f_{2}^{\dagger} f_{1}^{\pdag}$. In terms of these 
slave particles operators, the Hamiltonian for this system can be written as
\bn
H &=& \sum_{\eta, k}\epsilon _{\eta k} 
c_{\eta k }^{\dagger }c_{\eta k }^{\pdag} + \epsilon_{1} f_{1 }^{\dagger } f_{1 
}^{\pdag} + \epsilon_{2} f_{2 }^{\dagger } f_{2 }^{\pdag} \cr
&& + t_{c} (f_{1}^\dagger f_{2}^\pdag + f_{2}^\dagger f_{1}^\pdag)+ (2\epsilon_{d} +U) 
d^{\dagger} d \cr
&& + \sum_{k} [V_{L1} e^{i\varphi /4} c_{L k }^{\dagger } (e^{\dagger} f_{1 } + 
f_{2}^{\dagger} d) +{\rm {H.c.}}] \cr 
&& + \sum_{k} [V_{L2} e^{-i\varphi /4} c_{L k }^{\dagger } (e^{\dagger} f_{2 } + 
f_{1}^{\dagger} d) +{\rm {H.c.}}] \cr 
&& + \sum_{k} [V_{R1} e^{-i\varphi /4} c_{R k }^{\dagger } (e^{\dagger} f_{1} + 
f_{2}^{\dagger} d) +{\rm {H.c.}}] \cr
&& + \sum_{k} [V_{R2} e^{i\varphi /4} c_{R k }^{\dagger } (e^{\dagger} f_{2} + 
f_{1}^{\dagger} d) +{\rm {H.c.}}],
\label{hamiltonian2}
\en
where $c_{\eta k }^{\dagger}$ ($c_{\eta k }^{\dagger}$) are the creation (annihilation) 
operators for electrons with moment $k$, and energy $\epsilon_{\eta k}$ in the lead 
$\eta$ ($\eta=L,R$). $V_{\eta j}$ denotes the hopping matrix element between the dot and 
the lead and $\varphi\equiv 2\pi \Phi/\Phi_{0}$ accounts for the enclosed magnetic flux 
inside the AB interferometer ($\Phi_{0}=h/e$ is the flux quantum). $t_{c}$ is the 
interdot hopping coupling. 

We evaluate the statistical expectations of the rate of change of the density matrix 
elements $\rho_{ij}$ with the Hamiltonian (\ref{hamiltonian2}) and modified quantization 
Eq.\,(\ref{quan}). After tedious but straightforward calculations, we obtain
\begin{widetext}
\bn
\dot{\rho}_{00} &=& \langle i[H, e^{\dagger} e] \rangle = \frac{1}{2\pi} \sum_{ k} \big 
\{ [{\bf V}_{e}^{\dagger} {\bf G}_{k,e}^{<}(t,t)]_{11} - [{\bf G}_{e,k}^{<}(t,t) {\bf 
V}_{e}]_{11} + [{\bf V}_{e}^{\dagger} {\bf G}_{k,e}^{<}(t,t)]_{22} - [{\bf 
G}_{e,k}^{<}(t,t) {\bf V}_{e}]_{22}\big \}, \label{r00} \\
\dot{\rho}_{ii} &=& \langle i[H, f_{i}^{\dagger} f_{i}^{\pdag}] \rangle
= \frac{1}{2\pi} \sum_{ k} \big \{ [{\bf G}_{e,k}^{<}(t,t) {\bf V}_{e}]_{ii} - [{\bf 
V}_{e}^{\dagger} {\bf G}_{k,e}^{<}(t,t)]_{ii} + [{\bf V}_{d}^{\dagger} {\bf 
G}_{k,d}^{<}(t,t)]_{ii} - [{\bf G}_{d,k}^{<}(t,t) {\bf V}_{d}]_{ii}\big \} \cr
&&\hspace{2cm} + it_{c}(\rho_{i\bar{i}}-\rho_{\bar{i}i}) , \\
\dot {\rho}_{dd} &=& \langle i[H, d^{\dagger}d] \rangle 
= \frac{1}{2\pi} \sum_{ k} \big \{ [{\bf G}_{d,k}^{<}(t,t) {\bf V}_{d}]_{11} - [{\bf 
V}_{d}^{\dagger} {\bf G}_{k,d}^{<}(t,t)]_{11} + [{\bf G}_{d,k}^{<}(t,t) {\bf 
V}_{d}]_{22} - [{\bf V}_{d}^{\dagger} {\bf G}_{k,d}^{<}(t,t)]_{22}\big \}, \\
\dot{\rho}_{12} &=& \langle i[H, f_{2}^{\dagger} f_{1}^{\pdag}] \rangle
= \frac{1}{2\pi} \sum_{ k} \big \{ [{\bf G}_{e,k}^{<}(t,t) {\bf V}_{e}]_{12} - [{\bf 
V}_{e}^{\dagger} {\bf G}_{k,e}^{<}(t,t)]_{12} + [{\bf V}_{d}^{\dagger} {\bf 
G}_{k,d}^{<}(t,t)]_{21} - [{\bf G}_{d,k}^{<}(t,t) {\bf V}_{d}]_{21}\big \} \cr
&&\hspace{2cm} + i(\epsilon_2-\epsilon_1) \rho_{12} + it_{c} (\rho_{11}-\rho_{22}), 
\label{r12} 
\en
\end{widetext}
where the statistical expectations involve the Fourier transformations of the 
time-diagonal parts of the matrix correlation functions in $2\times 2$ space $[{\bf 
G}_{e,k}^{<}(t,t')]_{ij} \equiv i \langle c_{j k}^{\dagger}(t') e^{\dagger}(t) 
f_{i}^{\pdag}(t)\rangle$, $[{\bf G}_{d,k}^{<}(t,t')]_{ij} \equiv i \langle c_{j 
k}^{\dagger}(t') f_{i}^{\dagger}(t) d(t)\rangle$, $[{\bf G}_{k, e}^{<}(t,t')]_{ij} 
\equiv i\langle f_{j}^{\dagger}(t')e(t') c_{i k}^{\pdag}(t)\rangle$, and $[{\bf G}_{k, 
d}^{<}(t,t')]_{ij} \equiv i\langle d^{\dagger}(t') f_{j}^{\pdag}(t') c_{i 
k}^{\pdag}(t)\rangle$. With the help of the Langreth analytic continuation rules, 
\cite{Langreth} we can relate these hybrid Green's functions to the dressed Green's 
functions of the central region:
\bn
{\bf G}_{k,e/d}^{<}(t,t') &=& \int dt_1 [{\bf g}_{k}^{r}(t,t_1) {\bf V}_{e/d} {\bf 
G}_{e/d}^{<}(t_1,t') \cr
&& + {\bf g}_{k}^{<}(t,t_1) {\bf V}_{e/d} {\bf G}_{e/d}^{a}(t_1,t') ], \cr
{\bf G}_{e/d,k}^{<}(t,t') &=& \int dt_1 [{\bf G}_{e/d}^{r}(t,t_1) {\bf 
V}_{e/d}^{\dagger} {\bf g}_{k}^{<}(t_1,t') \cr
&& + {\bf G}_{e/d}^{<}(t,t_1) {\bf V}_{e/d}^{\dagger} {\bf g}_{k}^{a}(t_1,t')], 
\label{gf}
\en
where ${\bf V}_{e}$ and ${\bf V}_{d}$ are two $2\times 2$ matrixs of the hopping 
elements defined by
\bn
{\bf V}_{e} &=& \left ( 
\begin{array}{cc}
V_{L1} e^{i\varphi/4} & V_{L2} e^{-i\varphi/4} \\
V_{R1} e^{-i\varphi/4} & V_{R2} e^{i\varphi/4}
\end{array}
\right ), \cr
{\bf V}_{d} &=& \left ( 
\begin{array}{cc}
V_{L2} e^{-i\varphi/4} & V_{L1} e^{i\varphi/4} \\
V_{R2} e^{i\varphi/4} & V_{R1} e^{-i\varphi/4}
\end{array}
\right ),
\label{hopping}
\en
and $[{\bf g}_{k}^{r,a,<,>}(t,t')]_{ij}=\delta_{ij} g_{ik}^{r,a,<,>}(t,t')$ are the 
exact Green's functions of the $i$th lead without coupling to the device. These retarded 
(advanced) and lesser (greater) GFs for the central region are defined as: $G_{o ij 
}^{r(a)}(t,t')\equiv \pm i\theta (\pm t \mp t')\langle \{ O_{i}^{\pdag} (t), 
O_{j}^{\dagger}(t') \}\rangle$, 
$G_{o ij }^{<}(t,t')\equiv i\langle O_{j}^{\dagger}(t')O_{i}^{\pdag} (t)\rangle$ and 
$G_{o ij }^{>}(t,t')\equiv-i\langle O_{i}^{\pdag} (t)O_{j}^{\dagger}(t') \rangle$ with 
$O_{j}=e^{\dagger} f_{j}^{\pdag}$ if $o=e$ and $O_{j}=f_{j}^{\dagger} d$ if $o=d$. 

In the following derivation we perform a ``gradient expansion" of Eq.~(\ref{gf}), which 
is first introduced by Davies {\it et al}. to get the rate equation for resonant 
tunneling in the sequential regime.\cite{Davies} First define center-of-mass and 
relative times by $T=(t+t')/2$ and $\bar{t}=(t-t')/2$. Then we assume that functions 
vary rapidly in the relative time $\bar{t}$ but only slowly in the center-of-mass time 
$T$. Finally, we take a Fourier transform from $\bar{t}$ to $\omega$, and the GFs 
$G(t,t')$ in Eq.~(\ref{gf}) becomes $G(\omega,T)$. According to 
Ref.~\onlinecite{Davies}, the lowest-order gradient expansion is a good approximation 
for sequential resonant tunneling. Therefore, we just remain the first term in the 
gradient expansion of the GFs $G(\omega,T)$ in Eq.~(\ref{gf}) and substitute these GFs 
and the hopping matrixs ${\bf V}_{e/d}$ into Eqs.~(\ref{r00})-(\ref{r12}). It is noted 
that the equal time in Eqs.~(\ref{r00})-(\ref{r12}) means $\bar{t}=0$ or an integral 
over all $\omega$. Under the weak coupling (dot-lead tunneling and interdot hopping) 
assumption and slowly varying in time $T$, the GFs in the isolated two QDs system can be 
expressed in terms of spectrum representation.\cite{Dong,Ma} Inserting these GFs into 
the Fourier forms of Eqs.~(\ref{gf}), we then obtain the final quantum rate equations in 
the sequential tunneling regime. Because our interesting is focused on studying quantum 
dynamics and photo response of this interferometer at zero temperature and large bias 
voltage, we will not intend to give the general expressions but the interesting readers 
could refer to our recent paper Ref.~\onlinecite{Ma}. Finally, at zero temperature and 
large bias voltage, the quantum rate equations are written as:
\bn
\dot{\rho}_{00}&=& \Gamma_{R1} \rho_{11} + \Gamma_{R2} \rho_{22} - (\Gamma_{L1 } + 
\Gamma_{L2 }) \rho_{00} \cr
&& + [\Gamma_{R12} e^{-i\varphi/2} \rho_{12} + {\rm {H.c.}}], \label{rc0} \\
\dot{\rho}_{11}&=& \Gamma_{L1} \rho_{00} + \widetilde{\Gamma}_{R2} \rho_{dd} -  
(\Gamma_{R1} + \widetilde{\Gamma}_{L2}) \rho_{11} + it_{c} (\rho_{12}-\rho_{21}) \cr
&& - [\frac{1}{2} ( \Gamma_{R12} e^{-i\varphi/2} + \widetilde{\Gamma}_{L12} 
e^{i\varphi/2} ) \rho_{12} + {\rm {H.c.}}], \label{rc1} \\
\dot{\rho}_{22}&=& \Gamma_{L2} \rho_{00} + \widetilde{\Gamma}_{R1} \rho_{dd} - 
(\Gamma_{R2}+ \widetilde{\Gamma}_{L1}) \rho_{22} + it_{c} (\rho_{21}-\rho_{12}) \cr
&& - [\frac{1}{2} ( \Gamma_{R12} e^{-i\varphi/2} + \widetilde{\Gamma}_{L12} 
e^{i\varphi/2} ) \rho_{12} + {\rm {H.c.}}], \label{rc2} \\
\dot{\rho}_{dd} &=& \widetilde{\Gamma}_{L2} \rho_{11} + \widetilde{\Gamma}_{L1} 
\rho_{22} -  (\widetilde{\Gamma}_{R1} +  \widetilde{\Gamma}_{R2}) \rho_{dd} \cr
&& + [\widetilde{\Gamma}_{L12} e^{i\varphi/2} \rho_{12} + {\rm {H.c.}}], \label{rc3} \\
\dot{\rho}_{12} &=& i(\epsilon_2-\epsilon_1) \rho_{12} + it_{c}(\rho_{11}-\rho_{22})+ 
\Gamma_{L12} e^{-i\varphi/2} \rho_{00} \cr
&& + \widetilde{\Gamma}_{R12} e^{i\varphi/2} \rho_{dd} - \frac{1}{2} ( \Gamma_{R1} + 
\Gamma_{R2} + \widetilde{\Gamma}_{L1} + \widetilde{\Gamma}_{L2}) \rho_{12} \cr
&&  - \frac{1}{2} ( \Gamma_{R12} e^{i\varphi/2} + \widetilde{\Gamma}_{L12} 
e^{-i\varphi/2}) (\rho_{11} + \rho_{22}), \label{rc4}
\en
along with the normalization relation $\rho_{00}+ \rho_{11}+ \rho_{22} + \rho_{dd}=1$ 
and $\rho_{21}=\rho_{12}^{*}$, with the definitions $\Gamma_{\eta i}=2\pi \sum_{k} 
|V_{\eta i}|^2 \delta (\omega-\epsilon_{\eta k})$ denoting the strength of coupling 
between the $i$th QD level and the lead $\eta$. Namely, $\Gamma_{L i}$ ($\Gamma_{R i}$) 
here describes the tunneling rate of electrons in to (out from) the $i$th QD when the 
other QD is empty. On the contrary, $\widetilde {\Gamma}_{L i}$ ($\widetilde {\Gamma}_{R 
i}$) describes the tunneling rate of electrons in to (out from) the $i$th QDs, when the 
other QD is already occupied by an electron, revealing the modification of the 
corresponding rates due to the Coulomb repulsion between the two QDs. The interference 
in tunneling events through the different pathways are explicitly described by 
$\Gamma_{\eta ij}$ and $\widetilde{\Gamma}_{\eta ij}$ for the singly-occupied channel 
and doubly-occupied channel, respectively, with the definitions $\Gamma_{\eta ij}=2 \pi 
\sum_{k} V_{\eta i} V_{\eta j} \delta (\omega-\epsilon_{\eta k})$. These tunneling 
parameters are taken as constant under the wide band limit. In addition, the 
contribution of two leads is indeed negative to the nondiagonal density matrix element's 
dynamic equation, leading to damping of the quantum superposition. It is obvious that 
these damping terms are different from the seirs-coupled 
QDs.\cite{Stoof,Hazelzet,Gurvitz}

Actually, the similar equations have already been developed for this system by using 
other schemes.\cite{Jiang,Marquardt} Jiang and co-workers\cite{Jiang} applied the 
Gurvitz's wavefunction method\cite{Gurvitz} to derive the modified rate equations and 
studied the temporary dynamics. However, it should be noted that their equations are 
different from ours for the nondiagonal density matrix element Eq.~(\ref{rc4}). 
Marquardt and Bruder\cite{Marquardt} started from the von Neumann equation of the 
reduced density matrix and obtained the rate equations at a finite temperature. They 
studied the dephasing in sequential tunneling due to electron-phonon interaction for the 
similar device without interdot hopping. Their equations at zero temperature coincide 
with ours in absence of the interdot hopping.

The particle current $I_{\eta}$ flowing from the lead $\eta$ to the interferometer can 
be evaluated from the rate of change of the electron number operator 
$N_{\eta}(t)=\sum_{k} c_{\eta k}^{\dagger}(t) c_{\eta k}^{\pdag}(t)$ of the lead 
$\eta$:\cite{Meir}
\bq
{I}_{\eta}(t)= -\frac{e}{\hbar}\Big\langle \frac{d{N}_{\eta}}{dt}\Big\rangle=
-i\frac{e}{\hbar}\Big\langle \Big [H, \sum_{k} c_{\eta k }^{\dagger}(t)
c_{\eta k }(t)\Big ] \Big\rangle . \label{i}
\eq
Ultimately, the current $I_{L/R}$ can be expressed in terms of the GFs:
\bn
I_{L/R} &=& ie\int \frac{d\omega}{2\pi} \sum_{k} \big \{ {\bf G}_{k,e}^{<}(\omega) {\bf 
V}_{e}^{\dagger} - {\bf V}_{e}^{\pdag} {\bf G}_{e,k}^{<}(\omega) \cr
&& + {\bf G}_{k,d}^{<}(\omega) {\bf V}_{d}^{\dagger} - {\bf V}_{d}^{\pdag} {\bf 
G}_{d,k}^{<}(\omega) \big \}_{11/22}. \label{ii}  
\en
Under the weak coupling approximation, it becomes
\bn
I_{L}/e &=& - (\Gamma_{L1} + \Gamma_{L2}) \rho_{00} - \widetilde{\Gamma}_{L2} \rho_{11} 
- \widetilde{\Gamma}_{L1} \rho_{22} \cr
&& - \widetilde{\Gamma}_{L12} (e^{i\varphi/2} \rho_{12} + e^{-i\varphi/2} \rho_{21}). 
\label{iii}
\en 
The final term in the above expression comes from the contribution of the interference 
between the upper and lower pathways.

\section{Calculations and Discussions}

\subsection{Dc current and Quantum dynamics}

In this section, we first calculate the current through the parallel-coupled QDs in the 
presence of magnetic field and then study the quantum dynamical behavior of this system. 
The density of states and linear conductance of this system without the Coulomb 
interaction have been explored in detail in absence of the magnetic flux in 
literature.\cite{Shahbazyan,Ladron} It is found that when the system changes from a 
configuration in series to a completely symmetrical parallel geometry, the tunneling 
through the antibonding state is totally suppressed due to the perfect destructive 
quantum interference between the different pathways through the system.\cite{Ladron} 
With this point of view, we limit our investigation on the case of an asymmetry parallel 
configuration to guarantee that both the bonding and antibonding states have 
contribution to transport, in order to demonstrate the effect of interference on the 
photoresponse clearly. Here we assume the tunneling rates 
$\Gamma_{L1}=\Gamma_{R2}=\widetilde {\Gamma}_{L1}=\widetilde {\Gamma}_{R2}=\Gamma=0.2$ 
and $\Gamma_{L2}=\Gamma_{R1}=\widetilde {\Gamma}_{L2}=\widetilde {\Gamma}_{R1} =\Gamma' 
\leq 0.3\Gamma$. And we have $\Gamma_{L12}=\Gamma_{R12}=\sqrt{\Gamma \Gamma'}$. 

\begin{figure}[htb]
\includegraphics[height=1.8in,width=2.5in]{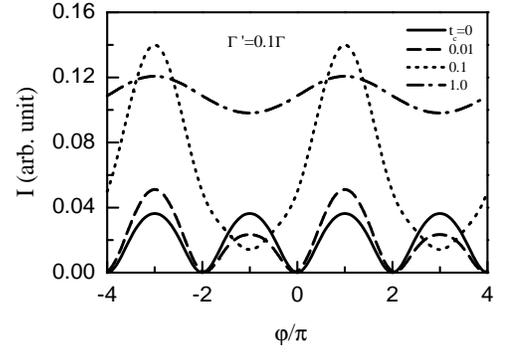}
\caption{The calculated stationary current $I$ as a function of the magnetic flux 
$\varphi$ for different interdot coupling $t=0$, $0.01$, $0.1$, and $1$ with 
$\Gamma=0.2$ and $\Gamma'=0.1\Gamma$.}
\label{fig2}
\end{figure}

Figure 2 displays the stationary current calculated from Eq.~(\ref{iii}) as a function 
of the renormalized magnetic flux $\varphi$ for different interdot couplings under the 
condition $\Gamma'=0.1 \Gamma$. It is clear that in the absence of the interdot coupling 
$t_{c}=0$, the current $I$ exhibits periodic oscillation with a period of $2\pi$. The 
current peaks appear at the phases of $(2n+1)\pi$ ($n$ is an integer number), and the 
current nearly vanishes at the phases of $2n \pi$. This is the main feature of the 
conventional AB effect. However, when the interdot coupling turns on, the periodicity of 
the AB oscillation becomes $4\pi$. In the new AB oscillation pattern, the first current 
peak also locates at the phase of $\pi$. But the positions of the current valleys move 
from the original phases of $0$ and $2\pi$ to the phases of $-\pi$ and $3\pi$. Similar 
characteristic has been also pointed out in Ref.~\onlinecite{Jiang}.    

\begin{figure}[htb]
\includegraphics[height=4.in,width=3.5in]{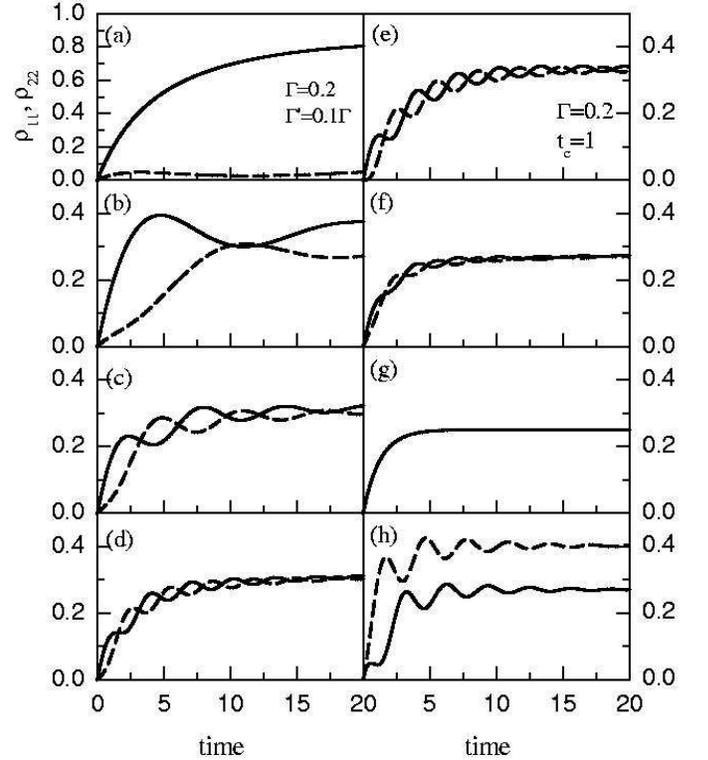}
\caption{The calculated time evolutions of the electron-occupation probabilities in the 
two QDs $\rho_{11}$ (solid lines) and $\rho_{22}$ (dashed lines). (a)-(d): Different 
interdot couplings $t_{c}=0$, $0.01$, $0.1$, and $1$ with $\Gamma'/\Gamma=0.1$; (e)-(g): 
Different ratios of the two tunneling rates $\Gamma'/\Gamma=0$, $0.5$, and $1$ with 
$t_{c}=1$; (h): $\Gamma'/\Gamma=1$ and $\varphi=\pi$. The time unit is $1/5\Gamma$.}
\label{fig3}
\end{figure}

We then study the quantum dynamical behavoir of the system. In calculation, we assume 
that the system initially occupies the empty state $\rho_{00}=1$. Figure 3 shows the 
time evolutions of the electron probability densities $\rho_{11}$ and $\rho_{22}$ for 
the states $|1\rangle_{1} |0\rangle_{2}$ and $|0\rangle_{1} |1\rangle_{2}$. In the 
absence of the interdot hopping $t_{c}=0$, the probability $\rho_{11}$ is nearly equal 
to $1$, while $\rho_{22}$ is nearly $0$ after a long time, because the higher 
``injection" rate to the first QD makes electrons occupy this QD, and the higher 
``escape" rate of the second QD leads to no electrons staying at all [Fig.~3(a)]. When 
we turn on the interdot hopping $t_{c}$, the probability in the second QD $\rho_{22}$ is 
largely enhanced even for very small interdot hopping $t_{c}=0.01$ as depicted in 
Fig.~3(b). It is also shown that these probabilities display temporal oscillations at 
small time, meaning that the electron vibrates back and forth between the two QD. And 
the oscillation period is shortened by raising the interdot hopping $t_{c}$. This is the 
well-known small-$t$ oscillation behavior in the two-level system: $\rho_{11}\sim 
\cos^2(t_c t)$ and $\rho_{22}\sim \sin^2(t_c t)$. However, we will find out some new 
characteristics in this small-$t$ oscillation for the parallel-coupled QDs. As indicated 
in Fig.~3(d)-(f), the raising tunneling rate of the additional pathway $\Gamma'$ 
gradually destroys the oscillations of the two probabilities due to the increasing 
quantum interference between the additional pathway and the original one. For the equal 
tunneling rates of the two pathways, $\rho_{11}$ and $\rho_{22}$ are of course equal and 
show no oscillations [Fig.~3(g)]. This is a consequence of perfect destructive quantum 
interference between the different pathways through the parallel-coupled QDs. 
Furthermore, it is already known that applying the magnetic field will change the 
scattering phase shift in every QDs, and then vary the interference patterns. As a 
result, the totally vanishing small-$t$ oscillations can be recovered to some extent by 
threading a magnetic flux, which can be observed in Fig.~3(h) for a normalized magnetic 
flux $\varphi=\pi$.                      

\subsection{Photoresponse}

In this subsection, we study the time-dependent tunneling through the AB interferometer 
in the nonlinear regimes. We assume that a time-dependent oscillation signal is applied 
to the two interacting QDs, so that the bare energy detuning becomes time-dependent 
$\epsilon_2 - \epsilon_1 = \epsilon_0 + \delta \cos \Omega t$, where $\delta$ is the 
amplitude and $\Omega$ the frequency of the externally applied signal. In the following, 
we use the approach developed by Stoof and Nazarov\cite{Stoof} to investigate the 
limiting case of a slight amplitude, $\delta \ll \Omega, \Gamma_{L/R}$, the linear 
photoresponse.

For simplicity, we assume the interdot Coulomb interaction $U$ is infinite, whereas only 
one electron can be found inside the system, so $\rho_{dd}=0$ and all 
$\widetilde{\Gamma}$ are equal to $0$. The quantum rate equations 
(\ref{rc1})-(\ref{rc4}) simplify to
\bn
\dot{\rho}_{11} &=& \Gamma \rho_{00} - \Gamma' \rho_{11} +it_{c}(\rho_{12} - \rho_{21}) 
\cr
&& - \frac{1}{2} [\Gamma_{R12} e^{-i\varphi/2} \rho_{12} + {\rm H.c.}] , \label{erc1} \\
\dot{\rho}_{22} &=& \Gamma' \rho_{00} - \Gamma \rho_{22}+ it_{c} (\rho_{21}-\rho_{12}) 
\cr
&& - \frac{1}{2} [\Gamma_{R12} e^{-i\varphi/2} \rho_{12} + {\rm H.c.}] , \\
\dot{\rho}_{12} &=& i(\epsilon_2-\epsilon_1) \rho_{12} + it_{c} (\rho_{11} - \rho_{22}) 
+ \Gamma_{L12} e^{-i\varphi/2} \rho_{00} \cr
&& - \frac{1}{2} \Gamma_{R12} e^{i\varphi/2} (\rho_{11} + \rho_{22}) - \frac{1}{2}( 
\Gamma' + \Gamma) \rho_{12} , \cr
&& \label{erc3}
\en
with $\rho_{00}+ \rho_{11}+ \rho_{22}=1$. Correspondingly, the stationary current 
simplifies as:
\bq
I_{L}/e = - (\Gamma + \Gamma') (1-\rho_{11}-\rho_{22}). \label{i0}
\eq
We rewrite the simplified quantum rate equations in matrix notation:
\bq
\frac{\partial {\bm \rho}}{\partial t}=({\bm \Gamma}+ {\bf T}+ {\bm \epsilon}_0 + {\bm 
\delta} \cos \Omega t){\bm \rho} + {\bm c},
\eq
where ${\bm \rho}=(\rho_{11}, \rho_{22}, \rho_{12}, \rho_{21})^{T}$, ${\bm c}=[\Gamma, 
\Gamma', \Gamma_{L12} e^{-i\varphi/2}, \Gamma_{L12} e^{i\varphi/2}]^{T}$, and ${\bm 
\Gamma}$, ${\bf T}$, ${\bm \epsilon}_0$, and ${\bm \delta}$ are the matrix forms of 
tunneling rates, hopping between dots, time-independent and time-dependent energy 
differences corresponding to Eqs.~(\ref{erc1})-(\ref{erc3}). The stationary solution of 
these equations without irradiation is expressed as: 
\bq
{\bm \rho}^{(0)}=-({\bm \Gamma} + {\bf T} + {\bm \epsilon}_0)^{-1} {\bm c}. \label{rho0}
\eq
Under the condition of small oscillation amplitude, the time-dependent density matrix 
elements can be expanded as:
\bq
{\bm \rho}={\bm \rho}^{(0)} + {\bm \rho}^{(1+)} e^{i\Omega t} + {\bm \rho}^{(1-)} e^{-i 
\Omega t} + {\bm \rho}^{(2)},
\eq
where ${\bm \rho}^{(1\pm)}$ and ${\bm \rho}^{(2)}$ are the positive (negative) frequency 
part of the first order correction and frequency-independent second order correction to 
the stationary solution ${\bm \rho}^{(0)}$, respectively. They are proportional to the 
small amplitude of the oscillating signal $\delta$. Substituting the above equation into 
the time-dependent rate equations and expanding according to the perturbation parameter 
$\delta$, therefore, we obtain
\bn
{\bm \rho}^{(1\pm)} &=& -\frac{1}{2}({\bm \Gamma}+ {\bf T}+ {\bm \epsilon}_0 \mp i\Omega 
{\bm I})^{-1} {\bm \delta} {\bm \rho}^{(0)}, \\
{\bm \rho}^{(2)} &=& -\frac{1}{2} ({\bm \Gamma}+ {\bf T}+ {\bm \epsilon}_0)^{-1} {\bm 
\delta} ({\bm \rho}^{(1+)} + {\bm \rho}^{(1-)}),
\en
${\bm I}$ being the unit matrix. The first order correction ${\bm \rho}^{(1\pm)}$ 
provides oscillatory terms and has no contribution to the dc current (time average 
current) Eq.~(\ref{i0}). The remaining lowest order contribution of the oscillating 
signal to the dc current comes from the second order correction ${\bm \rho}^{(2)}$. It 
is called photocurrent $I_{ph}$:
\bq
I_{ph}= (\Gamma + \Gamma') [{\bm \rho}_{11}^{(2)} + {\bm \rho}_{22}^{(2)}].
\eq

\begin{figure*}[htb]
\begin{center}
\makebox{\includegraphics[height=3.in,width=3.5in]{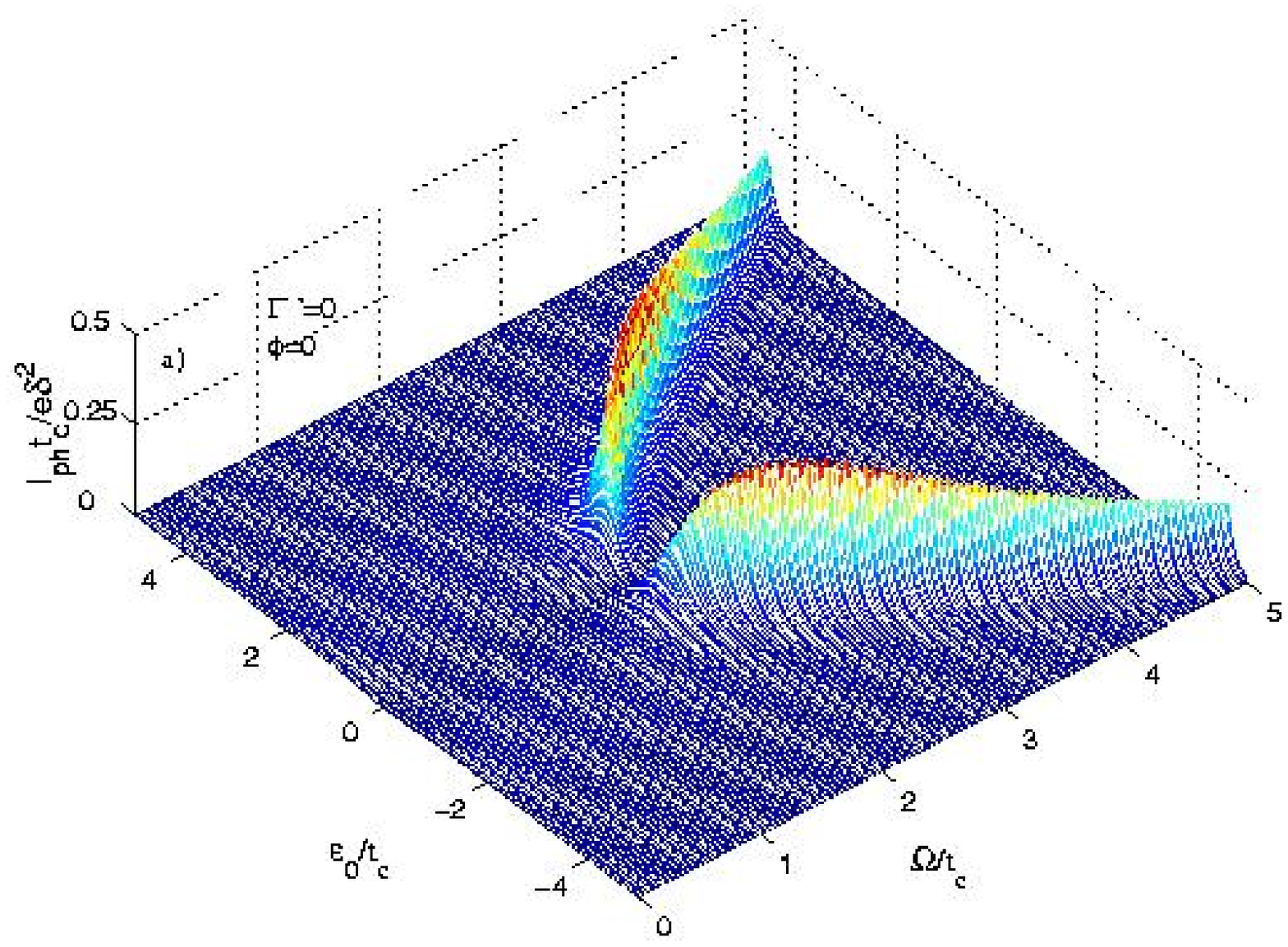}
         \includegraphics[height=3.in,width=3.5in]{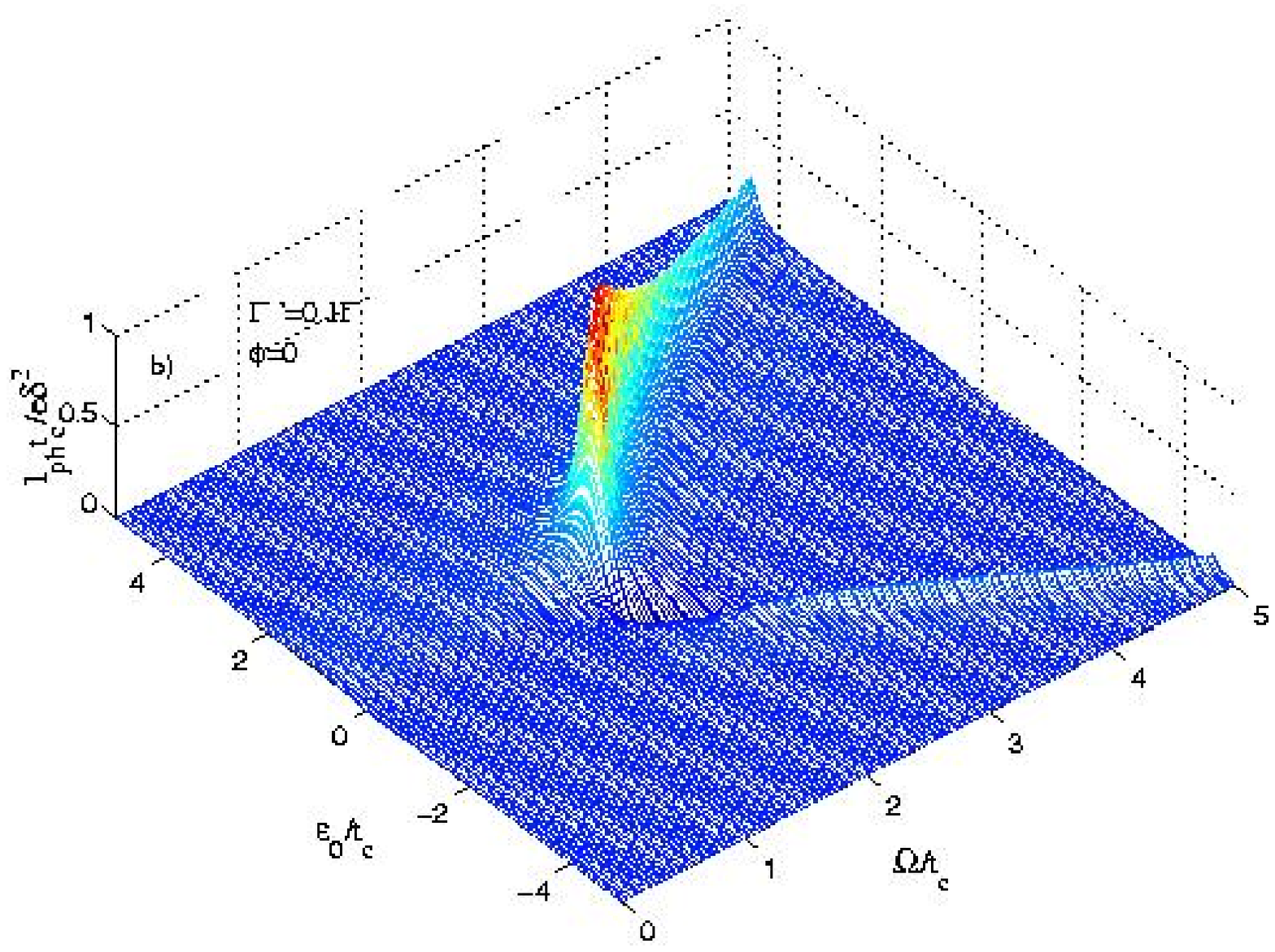} }\par
		 \vspace*{1cm}
\makebox{\includegraphics[height=3.in,width=3.5in]{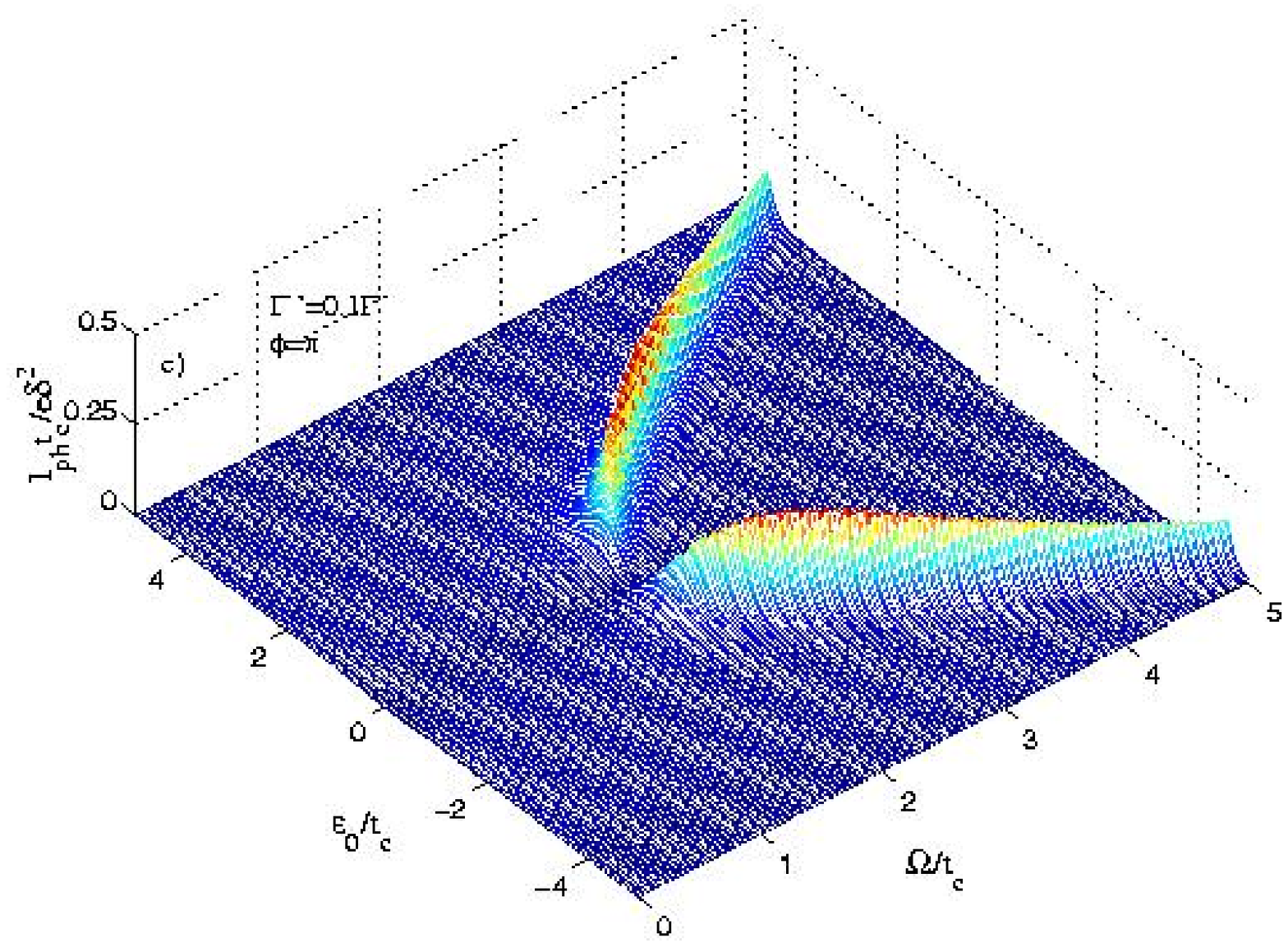}
	     \includegraphics[height=3.in,width=3.5in]{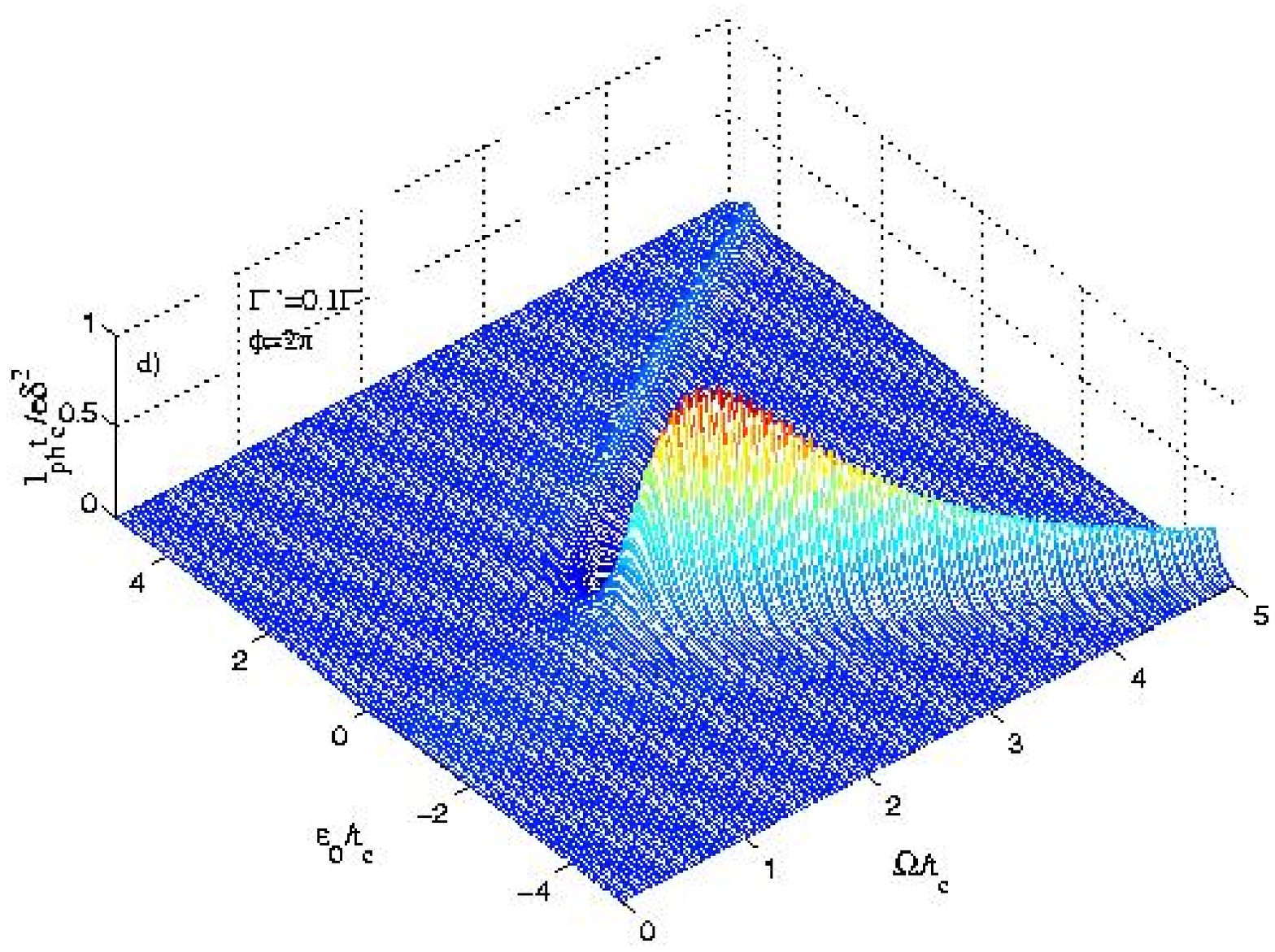}}
\caption{The normalized photocurrent of the parallel-coupled QDs, as a function of the 
bare level difference $\epsilon_0/t_c$ between the two QDs and the frequency 
$\Omega/t_c$ of the irradiation. (a): The rate $\Gamma'=0$ and the renormalized magnetic 
flux $\varphi=0$; (b)-(c): $\Gamma'/\Gamma=0.1$, the magnetic fluxes $\varphi=0$, $\pi$, 
and $2\pi$.}\label{fig4}
\end{center}
\end{figure*}

In Fig.~4 we plot the calculated photoresponses stationary current as a function of 
$\epsilon_0$ and the irradiation frequency $\Omega$ for a given interdot hopping 
$t_{c}=1$. Fig.~4(a) displays the result without the additional pathway $\Gamma'=0$, 
which is just the coupled QDs in series. So the characteristic of this figure is the 
same as Fig.~2 in Ref.\onlinecite{Stoof}: 1) Two branches of resonant satellite peaks 
for $\epsilon_0$ and $\Omega$ satisfying $\Omega^2=\epsilon_0^2+4 t_{c}^2$ (positive 
branch for $\epsilon_0>0$, negative branch for $\epsilon_0<0$), i.e., resonant PAT 
occurs when the emission or absorption of one photon can match the renormalized energy 
difference $\sqrt{\epsilon_0^2+4 t_{c}^2}$ of the two levels; 2) No satellite peaks 
appearance for the case of frequencies below $2t_{c}$. This is because the condition of 
resonant PAT is never satisfied due to the lower energy $\hbar\Omega$ of the applied 
irradiation. They are the two general conditions for occurrence of the resonant PAT. 
However, when we switch on the additional pathway with a small rate 
$\Gamma'/\Gamma=0.1$, we can observe that the photocurrent behavior is significantly 
altered as depicted in Fig.~4(b)-(c), although the two main characteristics above 
mentioned keep unchanged. It is clear in Fig.~4(b) that the positive branch of the PAT 
peaks is enhanced but the negative branch is suppressed nearly to zero amplitude. We 
attribute this phenomenon to the interference between the two pathways electrons can 
travel through in this new configuration. The scattering phase shifts are different when 
electrons pass through the two QDs with different bare energy levels. Consequently, the 
positive energy spacing $\epsilon_0>0$ leads to constructive quantum interference. On 
the contrary, the negative energy difference $\epsilon_0<0$ induces destructive quantum 
interference. This is the reason of the new resonant PAT pattern shown in Fig.~4(b). 
This explanation can be further substantiated by the fact that the  application of the 
magnetic fluxes will change the interference fashion, and thus modify the photocurrent 
behavior. In Fig.~4(c) the calculated photocurrent is plotted for a given magnetic flux 
$\varphi=\pi$. An amazing finding is that the photocurrent becomes very similar to the 
result of the series-coupled QDs, except with a reduced magnitude. Recalling that the 
periodicity of the AB oscillation is $4\pi$ in the presence of nonzero interdot hopping 
as pointed out in the above subsection, it is easy to imagine that we will obtain the 
opposite photoresponse to the situation depicted in Fig.~4(b) if the renormalized 
magnetic flux is set to be $2\pi$. The numerical results are plotted in Fig.~4(d). It is 
obvious that the negative branch of the resonant PAT peaks rises but the positive one 
declines.                              

\begin{figure}[htb]
\includegraphics[height=5in,width=2.5in]{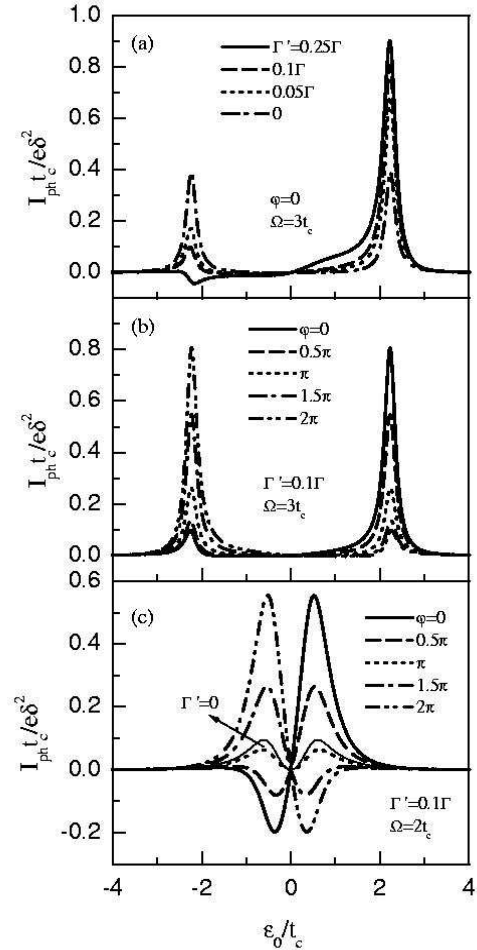}
\caption{Evolution of the two satellite peaks for different ratios $\Gamma'/\Gamma$ and 
different renormalized magnetic fluxes $\varphi$. (a): The plots are for 
$\Gamma'/\Gamma=0$, $0.05$, $0.1$, and $0.25$, respectively, at $\varphi=0$ and 
$\Omega=3t_{c}$; (b): The plots are for $\varphi/\pi=0$, $0.5$, $1.0$, $1.5$, and $2.0$ 
at $\Gamma'/\Gamma=0.1$ and $\Omega=3t_{c}$; (c): The plots are the same as (b) but for 
$\Omega=2t_{c}$. For the sake of comparison, the result of $\Gamma'=0$ is denoted as the 
thin line in (c).}
\label{fig5}
\end{figure}

In the following, we study in detail what actually happens for the reduced satellite 
peaks in the photoresponse stationary current by analyzing the evolution of the resonant 
PAT peaks as a function of the energy difference $\epsilon_0$ for different ratios of 
rates $\Gamma'/\Gamma$ and renormalized magnetic fluxes $\varphi$ at a given frequency 
of the applied irradiation (it must be bigger than or equal to $2t_{c}$). We plot the 
calculated results in Fig.~5. Figure 5(a) is for the frequency of $\Omega=3t_{c}$ and 
several rate ratios in the absence of magnetic field. We can observe more clearly that 
the two PAT peaks are always located at the points $\epsilon_{\pm}=\pm 
\sqrt{\Omega^2-4t_{c}^2}$ irrespective of whether the additional pathway is switched on 
or not. As the rate of the additional pathway is rising, the peak located at positive 
position $\epsilon_{+}$ increasingly heightens and the $\epsilon_{-}$ peak decreases. 
But the shapes of peaks remain similar with those of $\Gamma'=0$, in which case the two 
peaks are entirely identical and have a Lorentzian line shape near the resonance point 
$\epsilon_{\pm}$ as revealed in Ref.\onlinecite{Stoof}, until a stronger rate of the 
additional pathway $\Gamma'/\Gamma=0.25$. For this maximum rate considered here, the 
photocurrent displays as a negative dip with very small amplitude instead of a peak. It 
can be approximated by a Fano line shape. This behavior has a similar interpretation 
with those of the linear conductance and density of states when the configuration 
changes from series to parallel.\cite{Ladron} The interfering of two tunneling channels 
opens prominent perspectives for the Fano effect, i.e., asymmetric line shape of current 
at resonant point. In addition, the smaller frequency of irradiation leads to the 
appearance of the dip in photocurrent easier (at a smaller rate $\Gamma'$). Therefore, 
the Fano line shape can be observed more clearly at $\Omega=2t_{c}$ as shown in 
Fig.~5(c). 

The effect of the magnetic field on the photoresponse can be clearly observed from 
Fig.~5(b) and (c). When the renormalized magnetic flux $\varphi$ is equal to $\pi$, the 
Fano line shape of the resonant PAT peak absolutely changes back to the conventional 
Lorentzian shape if the resonant condition is satisfied. Finally, we conclude that for 
the moderate ratios of rates $\Gamma'/\Gamma$ and frequencies of irradiation, the 
photoresponse of the considered system expresses itself from the PAT Lorentzian peak to 
the PAT Fano peak by tuning the magnetic field threaded inside this interferometer.                  

\section{Conclusion}

In summary, we have presented the quantum rate equations in the sequential tunneling 
regime by means of the nonequilibrium Green's function for a mesoscopic AB ring with two 
tunneling-coupled QDs embedded in the two arms. Employing this set of quantum rate 
equations, we calculated the AB oscillation current, the temporal evolution of the 
electron-occupation probabilities in the two QDs, and the dc photocurrent as the 
photoresponse in the presence of a weak irradiation, at zero temperature and large bias 
voltage between the source and the drain.

Our numerical studies show that the permission of hopping between the two dots changes 
the AB oscillation period to $4\pi$ in comparison to the conventional $2\pi$ period 
oscillation observed in the typical AB effect. On the other hand, we find that the 
small-$t$ oscillation of the electron-occupation probabilities is also established in 
the two QDs, with a damping amplitude controlled by the asymmetry of the parallel 
configuration. When the configuration is completely symmetrical the small-$t$ 
oscillation is totally destroyed. Interestingly, this oscillation behavior is reobtained 
if we vary the enclosed magnetic flux due to the interference effect.

Finally, we have evaluated in detail the dc transport through the parallel-coupled QDs 
subject to a small external harmonic irradiation. It is found that, like in the 
series-coupled QDs, the photocurrent of this system exhibits extra resonant peaks when 
the frequency of the external signal matches the energy difference between the discrete 
states. Moreover, one branch of the PAT peaks in photocurrent is enhanced, while another 
branch is suppressed, which is dependent on the enclosed magnetic flux. This behavior is 
a consequence of quantum interference between the different pathways electrons can pass 
through, and does not exist for series-coupled QDs, where only one pathway exists. For 
some appropriate rate ratios $\Gamma'/\Gamma$ and frequencies of signal $\Omega$, the 
PAT peaks of photcurrent are composed of Lorentzian and Fano line shapes at the resonant 
points, respectively, which can also be controlled by applying magnetic fields.

In this paper, we study the time-dependent resonant tunneling through double QDs in a 
parallel configuration. This device has received wide attention in theoretical and 
experimental investigations at present, because its tunability makes it possible for 
further applications in quantum computation and quantum information. We hope our 
theoretical results about the novel PAT properties can stimulate experimental studies on 
this particular problem.                    

\begin{acknowledgments} 

B. Dong, I. Djuric and H. L. Cui are supported by the DURINT Program administered by the 
US Army Research Office. X. L. Lei is supported by Major Projects of National Natural 
Science Foundation of China (10390162 and 90103027), the Special Founds for Major State 
Basic Research Project (G20000683) and the Shanghai Municipal Commission of Science and 
Technology (03DJ14003).

\end{acknowledgments}

\end{document}